\documentclass[twocolumn]{aastex63}

\newcommand\methyl{CH$_3$CN}
\newcommand\methanol{CH$_3$OH}
\usepackage{gensymb}
\usepackage{comment}
\let\tablenum\relax
\usepackage{siunitx}
\usepackage[ampersand]{easylist}
\ListProperties(Hide=100, Hang=true, Progressive=3ex, Style*=$\bullet$ ,Style2*=$\circ$ ,Style3*=-- ,Style4*=\tiny$\blacksquare$ )

\received{XXXX}
\revised{YYYY}
\accepted{ZZZZ}
\submitjournal{ApJ}

\shorttitle{\methyl{} abundance gradients in HOPS-370}
\shortauthors{Walls, L.G. et al.}
\graphicspath{{./}{}}
\begin{document}
\title{Chemical abundance gradients of organic molecules within a protostellar disk}

\correspondingauthor{Levi G. Walls}
\email{levigwal@umich.edu}

\author[0000-0001-6253-1630]{Levi G. Walls}
\affiliation{Department of Astronomy, University of Michigan, Ann Arbor, MI 48109-1107}

\author[0000-0002-2555-9869]{Merel L.R. van 't Hoff}
\affiliation{Department of Astronomy, University of Michigan, Ann Arbor, MI 48109-1107}

%\collaboration{1}{(AAS Journals Data Scientists collaboration)}

\author[0000-0003-4179-6394]{Edwin A. Bergin}
\affiliation{Department of Astronomy, University of Michigan, Ann Arbor, MI 48109-1107}
%\nocollaboration{1}



\begin{abstract}
Observations of low-mass protostellar systems show evidence of rich complex organic chemistry. Their low luminosity, however, makes determining abundance distributions of complex organic molecules (COMs) within the water snowline challenging. However, the excitation conditions sampled by  differing molecular distributions may  produce substantive changes in the resulting emission.  Thus, molecular excitation may recover spatial information from spatially unresolved data. By analyzing spatially-unresolved NOrthern Extended Millimeter Array (NOEMA) observations of CH$_3$OH and CH$_3$CN, we aim to determine if CH$_3$OH and CH$_3$CN are distributed differently in the protostellar disk around HOPS-370, a highly-luminous intermediate mass protostar. Rotational diagram analysis of CH$_3$OH and CH$_3$CN yields rotational temperatures of $198 \pm 1.2$ K and $448 \pm 19$ K, respectively, suggesting the two molecules have different spatial distributions. Source-specific 3D LTE radiative transfer models are used to constrain the spatial distribution of CH$_3$OH and CH$_3$CN within the disk. A uniform distribution with an abundance of $4\times10^{-8}$ reproduces the CH$_3$OH observations. In contrast, the spatial distribution of CH$_3$CN needs to be either more compact (within $\sim120$ au versus $\sim240$ au for CH$_3$OH) or exhibiting a factor of $\gtrsim 15$ increase in abundance in the inner $\sim55$ au.  A possible explanation for the difference in spatial abundance distributions of CH$_3$OH and CH$_3$CN is carbon-grain sublimation.
\end{abstract}



\section{Introduction}
Complex organic molecules (COMs) are carbon-bearing molecules with 6 or more atoms \citep{Herbst_2009}. Many COMs are formed in ice mantles on dust grains. For example, hydrogenation of CO ice can lead to the formation of simple oxygen-bearing COMs, such as methanol (CH$_3$OH), up to biologically relevant molecules like glycerol \citep{Watanabe_2003,Fuchs_2009,Oberg_2009,Fedoseev_2017}. Whether a COM will remain in the ice or transitions into the gas depends on energetics and the local temperature \citep[see, e.g., the review by][]{Minnisale_2022}. The binding energies of many COMs are similar to that of water, resulting in the transition between gas and ice to occur roughly around the water snowline ($\sim$ 150 K for pressures $P \sim 10^{-12}$ bar).  For example, the binding energy is 5640 K for amorphous solid water \citep{Bolina_2005a, Ulbricht_2006, Dulieu_2013}, 5728 K for methanol \citep{Bolina_2005b, Ulbricht_2006, Doronin_2015}, and 4954 K for methyl cyanide  \citep{Abdulgalil_2013, Bertin_2017}, as measured for carbonaceous grains.   

Based on this picture, the spatial distribution of COMs in circumstellar material is expected to be uniform post-sublimation within the snowline in the absence of gas phase chemical processing. However, for some species, for example, the nitrogen-bearing COM methyl cyanide, CH$_3$CN, (additional) gas phase formation routes have been proposed \citep{Hebrard_2009, Loison_2014, Walsh_2014, Wakelam_2015, Garrod_2022}. In addition, chemical processing in hot ($\gtrsim$300-500 K) gas may be possible, potentially following the destruction of carbonaceous grains \citep{Lee_2010, Anderson_2017, Klarmann_2018, Wei_2019,vanthoff20, Binkert_2023} or through high-temperature pathways that become active \citep{Garrod_2022}. 

For low-mass sources, the snowline lies within a few au from the star in disks up to a few tens of au from the star in protostellar envelopes, making resolving the snowline location technically challenging. As such, most studies of low-mass protostellar sources have focused on the detection of COMs \citep[e.g.,]{Oberg_2011,Belloche_2020,Yang_2021} or the chemical inventory of individual sources \citep[e.g.,][]{vanDishoeck_1995,Cazaux_2003,Bottinelli_2004,Jorgenson_2016}, rather than their spatial distribution (see reviews by e.g., \citealt{Caselli_2012,Jorgensen_2020}). 

Evidence for chemical changes in hot gas have been found toward high-mass protostars. \cite{Blake_1987} is one of the first studies of the high mass star forming region Orion KL to indicate the observed kinematic and spatial difference in N-COM and O-COM abundances could be due to hot gas phase chemistry. More recent studies on the Orion KL region also show abundance enhancements of N-COMs relative to O-COMs \citep{Caselli_1993, Feng_2015, Crockett_2015, Gieser_2021, Busch_2022}.  
\cite{Nazari_2023} finds the ratio of column densities in hot gas relative to warm gas for N-COMs is approximately 5 times higher than for O-COMs. While all of these studies toward high-mass systems indicate resolved chemical segregation \citep[see also the literature review by ][showing enhancement in $\sim$40\% of sources]{vanthoff20} no systematic study of the spatial distribution of COMs within the snowline of low-mass sources has been conducted.

The disk around the highly-luminous protostar HOPS-370 \citep{Chini97, Reipurth99, Furlan16} also displays evidence of hot-gas chemistry based on observations with the NOrthern Extended Millimeter Array (NOEMA; van 't Hoff et al. accepted). Being an intermediate mass protostar \citep[$\sim2.5 M_\odot$;][]{Tobin20} with a luminosity of $\sim$314 $L_\odot$ \citep{Furlan16}, this system is warm, which pushes the water snowline out to $\sim225$ au. This source has been well-studied and well-modeled: ALMA observations resolved a disk structure in the dust continuum emission (of 62 au) and rotation in multiple molecular lines perpendicular to the outflow direction, suggesting the presence of a near edge-on (inclination of $74\degree$) gas disk \citep{Tobin20} that subtends a radial extent of $\sim 0\farcs5$. As such, HOPS-370 is a good target to study the spatial distribution of COMs within the snowline.

In this paper, we use spatially unresolved NOEMA observations of \methanol{} and \methyl{} in the disk around HOPS-370 to probe their spatial abundance distributions within the snowline using molecular excitation and source-specific radiative transfer modeling.
In \S~\ref{sec:probe}, we show how \methyl{} can be and is used as a temperature probe. 
In \S~\ref{subsec:obs} and \S~\ref{subsec:analysis}, we present the \methanol{} and \methyl{} observations toward HOPS-370 and our analysis that indicates differences in the spatial distributions of these species within the NOEMA beam.  These results are discussed in \S~\ref{sec:discussion} and the main conclusions are summarized in \S~\ref{sec:conclusion}.

\section{\methyl{} as a temperature probe} \label{sec:probe}
Symmetric-top molecules like \methyl{} are excellent probes for the gas temperature. These molecules break symmetry along the rotational axis in one dimension.  Thus their energy states are characterized by the two quantum numbers J (total angular momentum) and K (the projection of J onto the axis of symmetry).  
Radiative transitions for symmetric-tops have $\Delta$J = 1 and $\Delta$K = 0, while $\Delta \mathrm{K} \neq$ 1 transitions are forbidden. 
The selection rules for dipole transitions imply collisions dominate over radiative processes when relative populations are explored across K ladders (same J - different K), and thus probe the gas temperature \citep{Ho_1983, Loren_1984, Bergin_1994}.  To further enhance this capability, the rotational emissions of symmetric top molecules for transitions with the same J but different K are found closely spaced in frequency. Thus a single observation might capture multiple energy states at once.

In general, the excitation temperature, $T_{\mathrm{exc}}$, defines the relative population of upper level $u$ to lower level $l$ in the Boltzmann distribution:

\begin{equation}
\frac{N_u}{N_l} = \frac{g_u}{g_l} \exp{\left( -\frac{E_u - E_l}{k_B\ T_{\mathrm{exc}}}\right)}.
\end{equation}

\noindent If molecular emission is in LTE then the excitation temperature is equal to the kinetic temperature ($T_{\rm{k}}$).  For complex molecules, with multiple emission lines sampling a range of excitation states, it is common to use the rotation diagram method \citep{Goldsmith_1999} to measure the rotational temperature, $T_{\rm{rot}}$.  Here the rotational temperature is a relative measure of the excitation temperature.  Below we explore the relation of $T_{\rm{rot}}$ and $T_{\rm{k}}$ for the J=14-13 transition of CH$_3$CN using a series of statistical equilibrium calculations to lay the ground work for our subsequent observational analysis.  

The methodology laid out by \cite{Goldsmith_1999} states
\begin{equation} \label{eq:rotdiag}
\ln \left( \frac{N_u}{g_u} \right) = \ln \left( \frac{N_{T}}{Q(T_{\rm{rot}})} \right) - \frac{E_u}{T_{\rm{rot}}},
\end{equation}
where $N_u$ is the upper level column density of the molecule, $g_u$ the statistical weight of that level and $E_u$ is the upper energy level in units of K. The total column density of the molecular gas is given by $N_{T}$ and the partition function is $Q(T_{\rm{rot}})$. 

For optically thin transitions, $\ln (N_u / g_u)$ is related to the observed total integrated flux, $W$ in units of K km s$^{-1}$, via
\begin{equation}
\frac{N_u}{g_u} = \frac{8\pi k_B}{h\ c^3} \frac{\nu^2\ W}{A_{ul}\ g_{u}},
\end{equation}
where $k_B$ is the Boltzmann constant, $G$ is the gravitational constant, $h$ is Planck's constant, $c$ is the speed of light, $A_{ul}$ is the Einstein coefficient for spontaneous emission, and $\nu$ is the transition frequency.  

With Eq. \ref{eq:rotdiag} presented as such, a linear regression can be fit in semi-log space to the data, $\ln(N_u/g_u)$, as a function of upper energy level, $E_u$. The intercept is then proportional to the total column density and the negative slope yields the rotational temperature. For a fixed abundance, lowering the rotational temperature, $T_{\rm{rot}}$, steepens the slope.

\begin{figure}[ht!]
    \centering
    \includegraphics[width=\columnwidth]{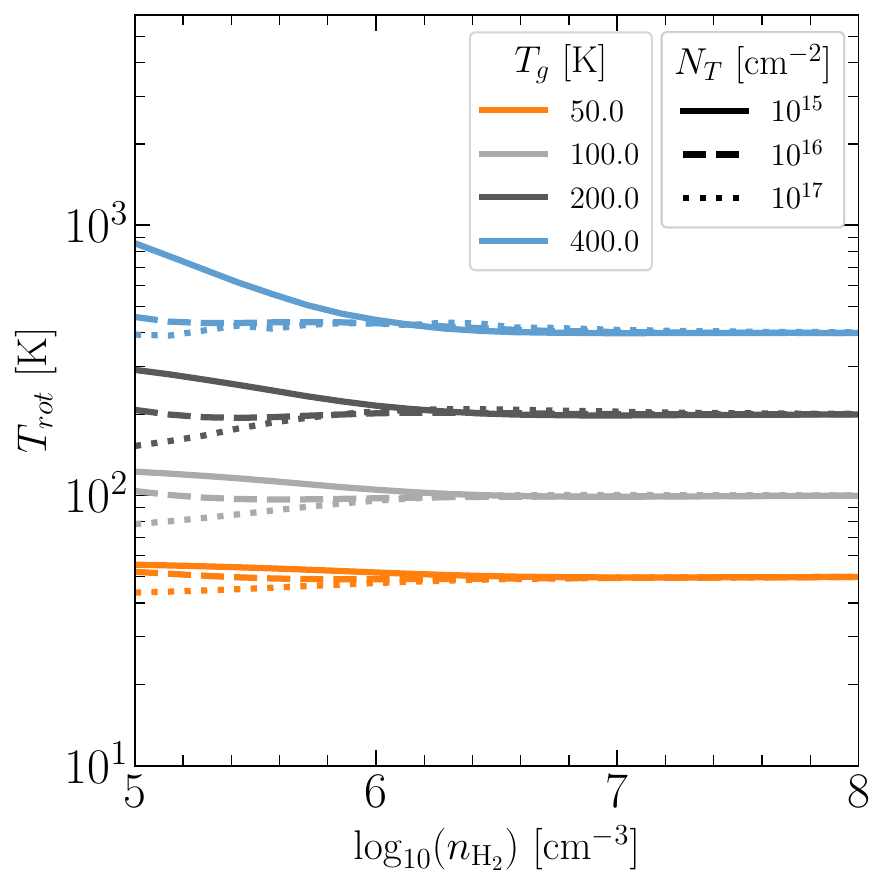}
    \caption{Rotational temperature versus gas density derived from \texttt{RADEX} simulations of J=$14_\mathrm{K} - 13_{\mathrm{K}}$, K=0-12 \methyl{} transitions using a rotation diagram analysis. Different colors indicate different gas (kinetic) temperatures and different linestyles indicate different total column densities for CH$_3$CN.  
    \label{fig:radex}}
\end{figure}

We use  \texttt{RADEX} slab model radiative transfer simulations \citep{radex} of \methyl{} J=$14_\mathrm{K} - 13_{\mathrm{K}}$, K=0-12 transitions for different kinetic (gas) temperatures, H$_2$ density, and total CH$_3$CN column densities.  The spectroscopic data and collisional rate coefficients we adopt in these simulations are provided by \cite{Green_1986,Mueller_2009,Mueller_2015}. For a given set of physical conditions \texttt{RADEX} calculates the line emission, from which upper level column densities and rotational temperatures are derived using the method outlined above.  Fig. \ref{fig:radex} displays the resulting rotational temperatures versus gas density for different kinetic (gas) temperatures, and total \methyl{}  column densities.   Based on this analysis, LTE (in this case $T_{\rm{rot}}$ = $T_{\rm{k}}$) is reached near densities of $n_{\mathrm{H_2}} \sim 10^6$ cm$^{-3}$.  Furthermore, LTE conditions are retrieved (within a factor of 2) when $n_{\mathrm{H_2}} < 10^6$~cm$^{-3}$ provided $N_T >10^{16}$~cm$^{-2}$ and the emission is optically thin. 

Additionally, a statistical equilibrium argument illustrates that LTE is appropriate for realistic disk temperatures and densities such as those described above (see \S~\ref{subsubsec:structure} for our disk model). Ignoring radiative processes, a system is in LTE when densities exceed the critical density, i.e. $n_{\mathrm{H_2}} \gg n_{cr} \equiv A_{ul}/C_{ul}$, where $C_{ul}$ is the collisional rate coefficient from the upper level to the lower level. \cite{Khalifa_2020} show updated collisional rate coefficients for E-\methyl{} ($\mathrm{K} = 3n \pm 1$ for $n \geq 0$) which are lower by at most a factor of 5 as compared with the rates from \cite{Green_1986}; A-\methyl{} ($\mathrm{K} = 3n$ for $n \geq 0$) is even closer to agreement. We thus approximate the most-up-to-date E-\methyl{} collisional rate cofficients of \cite{Khalifa_2020} at 140 K (roughly corresponding to the lowest temperature in our disk model). Calculating the critical density in this most conservative case, we find values do not exceed $\sim3.4\times10^7$ cm$^{-3}$ (occurring for $\mathrm{K}\leq 2$ transitions). This is approximately an order of magnitude lower than $2.2\times10^8$ cm$^{-3}$ (corresponding to the lowest gas number density in our disk model). Other locations in the disk are increasingly dense and hot and, thus, would only further solidify the fact the disk is in LTE. This effectively validates the use of CH$_3$CN to probe the gas temperature as demonstrated previously \citep{Loren_1984}.

In our observational analysis we will explore, and contrast, the rotational temperatures of CH$_3$OH to CH$_3$CN.  
CH$_3$OH is an asymmetric top molecule and its rotational temperature relationship to the gas kinetic temperature is more complex.  However, if its emission is in LTE then a comparison between measured rotational temperatures may be revealing.
For analysis of observations in the PErseus ALMA CHEmistry Survey (PEACHES), \cite{Yang_2021} performs 1D, isothermal non-LTE radiative transfer calculations of \methanol{}. They find that for the $( 5_{1,  4})$ $A^- - ( 4_{1,  3})$ $A^-$ \methanol{} transition ($\nu_t = 0$) at (243915.788 MHz) transition, the excitation temperature approached the kinetic temperature to within 10-30\% for gas density estimates appropriate for this type of sources (between $\sim4\times10^{9}$ and $10^{12}$ cm$^{-3}$). A similar analysis is done by \cite{Jorgenson_2016} in their study of IRAS 16293, albeit for a different frequency range, but with a similar conclusion. Furthermore, models of \cite{Goldsmith_1999} show LTE is appropriate for \methanol{} at densities around $\sim10^9$ cm$^{-3}$, albeit for a low kinetic temperature, $T_{\rm{k}} = 50$~K. We target the same frequency range as the PEACHES sources and the density for the HOPS-370 disk falls within the same range (see \S~\ref{subsubsec:structure}), suggesting LTE is a valid assumption for our modeling of the optically thin emission spectrum of \methanol{}.

 \section{Observations} \label{subsec:obs}
HOPS-370 was observed by the Northern Extended Millimeter Array (NOEMA) as a part of the Chemistry And Temperature: Searching with NOEMA for Destruction Of carbon GrainS project (CATS-N-DOGS; HOPS-370 project code: W20AF) four times from December 2020 to March 2021 (van 't Hoff et al. accepted). In this work, we use the first three observations, which were at 1 mm spanning the frequency ranges of 240.9--248.6 GHz and 256.4--264.1 GHz each with 2 MHz ($\Delta v = 2.3$ km s$^{-1}$ spectral resolution) and spent 3.0, 1.9, and 2.6 hours, on source in configuration 11C, 11C, and 10A, respectively.  

The data are calibrated using the CLIC package of the \texttt{GILDAS} software\footnote{\url{https://www.iram.fr/IRAMFR/GILDAS}}, using the most recent \texttt{GILDAS} version available at the time of data reduction (July 2021). For each observation the bandpass calibrator was 3C84. The phase calibration was done using J0542-0913 and J0458-020 for each observation. The flux was calibrated using LKHA101 for all observations except for the second, in which MWC349 was used. 

Imaging and cleaning was done using the \texttt{MAPPING} package of \texttt{GILDAS}. Due to the line richness of the spectrum, the data was imaged without continuum subtraction and no self-calibration was performed. The resulting beam size is  $3.37\arcsec \times 0.6\arcsec (9^\circ )$. The gas disk of HOPS-370 is $\sim$0.5\arcsec in size ($\sim200$ au in diameter at $\sim$400 pc; \citealt{Tobin19, Tobin20}), and thus unresolved in these observations. The noise level has been measured in a region off source in the image cube and is 16 mK in 2 MHz channels.

\begin{figure*}[ht!]
\centering
\includegraphics[width=0.75\textwidth]{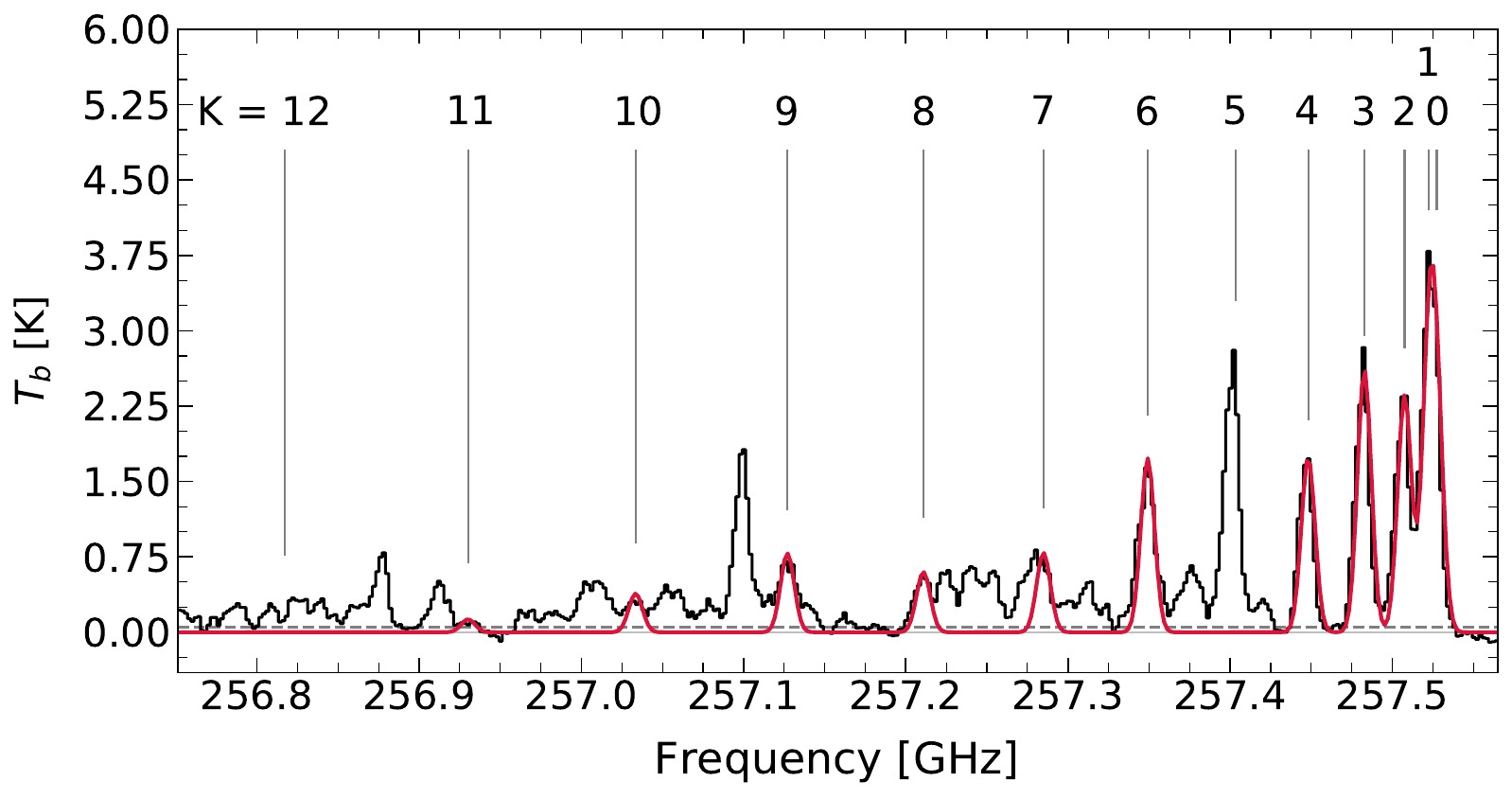}
\caption{NOEMA spectrum of the \methyl{} J=14-13 ladder extracted toward the continuum peak position of HOPS-370. Overlaid are Gaussian fits (crimson) to the observed transitions  (see also Table \ref{tab:ch3cn}). However, no fits are shown for the K = 5 or 12 lines (see text). The source velocity is $v_{LSR} = 11.2$ km s$^{-1}$ and each line has a full-width half-maximum of 11.5 km s$^{-1}$. All K-components are marked with vertical gray lines. The 1$\sigma$ noise level is 16 mK and is denoted by the dashed gray line.
\label{fig:noema_spectrum}}
\end{figure*}

After extracting spectra toward the continuum peak position, the continuum was subtracted using a statistical method as, for example, used by \citet{Jorgenson_2016} for the line-rich spectrum of the protostellar binary IRAS16293. This method entails the formation of a density distribution of flux values for the spectrum of each sideband. In the case of only continuum emission, the density distribution would represent a symmetric Gaussian centered at the continuum level. In the case of continuum and line emission, the Gaussian will contain an exponential tail toward higher values. The continuum level is then determined by fitting a skewed Gaussian to the distribution, and subtracted from the spectrum. The continuum-subtracted  \methyl{} $\mathrm{J} = 14_\mathrm{K} - 13_\mathrm{K}$ spectrum is shown in Fig. \ref{fig:noema_spectrum}. 

We securely detect 12 transitions of the $14_\mathrm{K} - 13_\mathrm{K}$ ladder of \methyl{} shown in Fig. \ref{fig:noema_spectrum} (see also Table \ref{tab:ch3cn}). The 16 mK noise level results in a weak detection of the K=11 \methyl{} transition. There are multiple other weak lines present near the K=12 frequency, which makes a detection of the K=12 transition very uncertain. We therefore do not report a detection of the K=12 transition. The K=5 \methyl{} transition at 257403.585 MHz coincides with the $(18_{3, 16})$ $A^+ - (18_{2, 17})$ $A^-$ \methanol{} transition ($\nu_t = 0$) at 257402.086 MHz. We thus report a detection of the K=5 \methyl{} transition, but do not include it (nor the \methanol{} transition) in the observed rotation diagram analysis. 

\section{Analysis} \label{subsec:analysis}
In this section, we model the emission of \methanol{} and \methyl{} via full 3D  radiative transfer calculations. The disk around HOPS-370 has a well-characterized physical structure \citep{Tobin20}. We use this structure for 3D radiative transfer modeling with the LIne Modeling Engine \citep[\texttt{LIME};][]{lime} which includes both gas and dust emission and their respective optical depth effects. We then convolve the predicted emission to the NOEMA beam (at the source distance) to match the spatially unresolved observations. Using the strategy outlined in Sect.~\ref{sec:probe}, we derive the rotational temperature for each molecular species.  A focus of our efforts will be to use the observed molecular excitation to set constraints on the overall spatial distribution of \methanol{} and \methyl{}. In \S~\ref{subsubsec:structure} we first describe the physical model in detail, and in \S~\ref{subsec:ch3oh} and \S~\ref{subsec:ch3cn_const} - \ref{subsec:ch3cn_jump} we present the results of our modeling efforts for \methanol{} and \methyl{}, respectively.

\subsection{Physical structure and model setup}  \label{subsubsec:structure}
Following the approach of \cite{Tobin20}, we adopt the gas disk structure  from their modelling results of \methanol{} emission, where 
the gas surface density, $\Sigma_{g}$, of the disk is assumed to be a power-law with an exponential cut-off \citep{LBP_1974}
\begin{equation} \label{eq:sigma}
\Sigma_{g}(r) = \Sigma_0 \Big( \frac{r}{r_c} \Big)^{-\gamma} \mathrm{exp}\Big[-\Big( \frac{r}{r_c} \Big)^{2 - \gamma}\Big], 
\end{equation}
where $\Sigma_0 = 51.25\ \mathrm{g\ cm^{-2}}$ is the gas surface density normalization, $r$ is the cylindrical radius, $r_c = 110$ au is the disk critical radius, and $\gamma = 0.89$ is the surface density power-law index. 

\begin{figure*}[ht!]
\centering
\includegraphics[width=\textwidth]{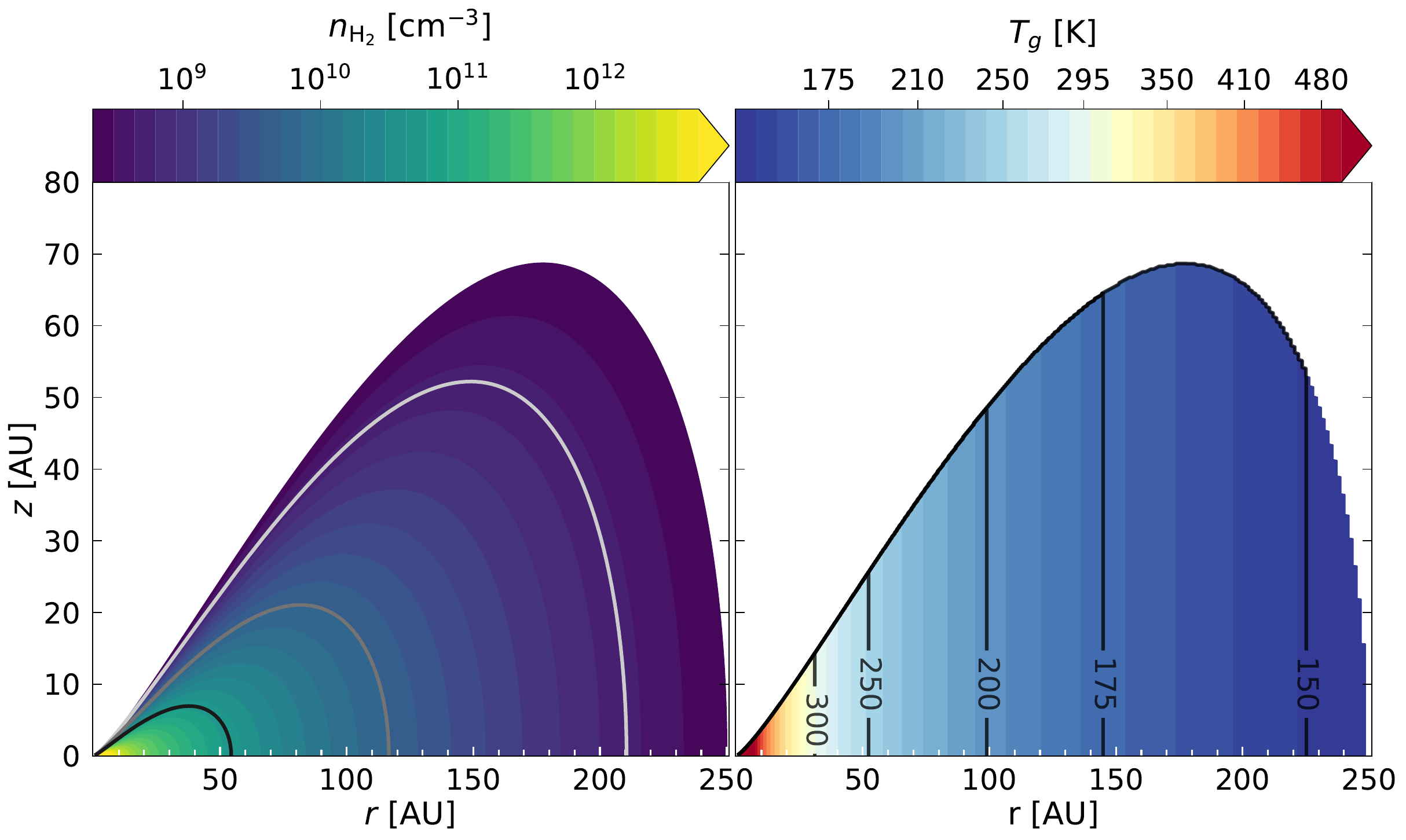}
\caption{H$_2$ number density (left) and gas temperature (right) in the adopted physical structure for the HOPS-370 disk based on modeling by \citet{Tobin20}. The contours in the left panel start at $5\times10^{8}$ cm$^{-3}$ (white) and increase by 1 dex per contour. The (vertically isothermal) contours in the right-hand panel show the gas temperature at different radii in the disk. \label{fig:dens_temp}}
\end{figure*}

The vertical structure of the disk is assumed to be in hydrostatic equilibrium, such that:
\begin{equation}    \label{eq:h}
h_{g}(r) = \Bigg(\frac{k_B\ r^3\ T_{g}(r)}{G\ M_*\ \mu_m m_\mathrm{H}} \Bigg)^{1/2}
\end{equation}
where $k_B$ is the Boltzmann constant, $G$ is the gravitational constant, $M_* = 2.37 M_\odot$ is the mass of the central protostar, $\mu_m = 2.37$ is the mean molecular weight \citep{Lodders_2003}, and $m_\mathrm{H}$ is the mass of the hydrogen atom. A radial power-law is adopted for the temperature profile, $T_{g}(r)$, 
\begin{equation} \label{eq:temp}
T_{g}(r) = T_0 \Big( \frac{r}{\mathrm{au}}\Big)^{-q}
\end{equation}
where $T_0 = 998.7$ K is the temperature at 1 au \citep{Tobin20}. The temperature power-law index $q=0.35$ was chosen based on results for other protostellar disks, e.g., L1527 \citep{Tobin_2013, merel_2018a}. 
With this temperature profile, as shown in the right panel of Fig. \ref{fig:dens_temp}, the 150 K water snowline is located at 225 au. 
See \S~\ref{subsec:iso} for more discussion on the effect temperature profile choice has on our modelling work.

The gas (mass) density then is found by integrating Eq. \ref{eq:sigma}:
\begin{equation}    \label{eq:rho}
\rho_{g}(r) = \frac{1}{\sqrt{2\pi}} \frac{\Sigma_{g}(r)}{h_{g}(r)} \exp \Bigg[-\frac{1}{2} \Big(\frac{z}{h_{g}(r)}\Big)^{2} \Bigg],
\end{equation}
where $\Sigma_{g}(r)$ is given by Eq. \ref{eq:sigma}, $h_{g}(r)$ is given by Eq. \ref{eq:h}, and $z$ is the cylindrical height above the midplane. The gas number density, $n_{\mathrm{H_2}}$, shown in the left panel of Fig. \ref{fig:dens_temp} is found by dividing the gas mass density (Eq. \ref{eq:rho}) by the mean molecular weight.

The dust density is assumed to be 1\% of the gas density \citep{Bohlin_1978}. Furthermore, the dust size distribution is assumed to follow $n(a) \propto a^{-3.5}$ \citep{MRN}, where $a$ is the radius of the dust, which is assumed to be between 0.005 and 1 $\mu$m. Dust grains are assumed to be mixtures of water ice (mass fraction: $1.19\times10^{-3}$), iron  (mass fraction: $2.53\times10^{-4}$), olivine (mass fraction: $2.51\times10^{-3}$), refractory and volatile organics (mass fraction: $4.13\times10^{-3}$), orthopyroxene (mass fraction: $7.33\times10^{-4}$), and troilite (mass fraction: $5.69\times10^{-4}$). Following the results of \citep{Tobin20}, we model the mass-weighted dust opacities presented in \cite{Pollack_1994} using \texttt{optool} \citep{optool}.

The main focus of this work is investigating the effects of varying the abundance distribution (e.g., constant abundance uniformly distributed over the entire disk) of \methanol{} and \methyl{}. The physical structure outlined above is used to perform LTE radiative transfer calculations for different transitions of \methanol{} and \methyl{} with \texttt{LIME} \citep{lime}. \texttt{LIME} solves the radiative transfer equation on a grid of Voronoi cells constructed from the Delauney triangulation of randomly chosen points, whose acceptance can be weighted based on, for example, a provided molecular abundance distribution. The result is an image cube containing the intensity for a prescribed spectral resolution. 

For both \methanol{} and \methyl{} we provide as input a Keplerian velocity profile and thermal Doppler broadening parameter in addition to the gas density (e.g. Eq. \ref{eq:rho}) and temperature structure (e.g. Eq. \ref{eq:temp}) as modeled by \cite{Tobin20}. Additionally we adopt in \texttt{LIME} the distance \citep[400 pc;][]{Kounkel_2018, Tobin_2020_2} and inclination angle \citep[74$\degree$;][]{Tobin20} for the disk around HOPS-370. The molecular abundance and its distribution for each molecular species is also provided and is the main quantity we vary.  

The spectroscopic information for both \methanol{} and \methyl{} (i.e. $\nu$, $g_u$, $E_u$, $A_{ul}$) is present in \texttt{CDMS} \citep{, Pickett_1998, cdms0, cdms1, cdms2} and provided in a file conforming to the \texttt{LAMDA} format \citep{lamda}. The spectroscopic data for \methyl{} is provided 
by \cite{Mueller_2009,Mueller_2015} (see also Table~\ref{tab:ch3cn}). 

Since \texttt{LIME} calculates the partition function internally using the energy levels in the provided molecular data file, we calculate the partition function in the same way for the analysis of the observations using 
\begin{equation}    \label{eq:partition_function}
Q(T) = \sum_i g_i \exp\left(-\frac{E_i}{k_B T}\right),
\end{equation}
where $g_i$ and $E_i$ is the statistical weight and energy of level, $i$, respectively. In calculating the total column density for each species, the partition function is evaluated at the rotational temperature. This is justified given LTE conditions have been established within the disk (see \S~\ref{sec:probe}).

Given the temperatures considered here, however, higher \methyl{} vibrational states can contribute to its partition function. We determine the multiplicative vibrational correction factor at the rotational temperature (a factor $\lesssim2$) by interpolating the correction factors provided in CDMS. Importantly, the derived rotational temperature does not depend on the partition function.

We include in our models the K=0-12 lines of the J=14$_\mathrm{K}$-13$_\mathrm{K}$ transition, each at a velocity resolution of 0.16 km s$^{-1}$. While we include the  K=0 and 1 lines in our observational and model rotation diagrams, we do not include them when performing the linear regression as these transitions are blended (see Table \ref{tab:ch3cn} for more information on what transitions are used in constructing the rotation diagrams in Figs.~\ref{fig:ch3cn_constrotdiag},~\ref{fig:ch3cn_comprotdiag},~and~\ref{fig:ch3cn_jumpcombrotdiag}).

The spectroscopic data for \methanol{} is provided by \cite{Xu_2008} (see also Table \ref{tab:ch3oh}).
From the \texttt{CDMS} entry for \methanol{}, which consists of  torsionally-excited $\nu_t \leq 2$ states up to J=40 with A- and E-symmetry, we construct a molecular data file containing 38,643 total transitions up to 3 THz. The Einstein $A_{ul}$ values are calculated as detailed by \cite{Pickett_1998}, and we calculate the partition function from the line list as detailed above. Each line is modeled with a velocity resolution of 1 km s$^{-1}$.

Each model yields an image cube containing the intensity in Jy/pixel for each line considered per molecule. For each line, the continuum is determined as the mean intensity in the first three velocity channels devoid of line emission. This is then subtracted from the rest of the velocity channels per line per molecule. This data is then converted to brightness temperature via the Rayleigh-Jeans law \citep{Kraus_1982}:
\begin{equation}
T_b = 1.222\times10^{3} \frac{I}{\nu^2\ \theta_{min}\ \theta_{max}},
\end{equation}
where $I$ is the spectral intensity in mJy, $\nu$ is the line frequency in GHz, and $\theta_{min} = 0\farcs6$ and $\theta_{max}=3\farcs37$ are the beam minor and major axes, respectively. After extracting the spectrum from the data cube, these modeled fluxes are then convolved with a 1D Gaussian assuming the velocity resolution of the NOEMA observations, i.e. $\Delta v = 2.3$ km s$^{-1}$ and integrated with respect to velocity. The models, in which abundances and spatial distributions for \methanol{} and \methyl{} are varied, are analyzed with rotation diagram analysis and, in this way, compared with observations. 

\subsection{\methanol{}} \label{subsec:ch3oh}
\begin{figure}
\includegraphics[width=\columnwidth]{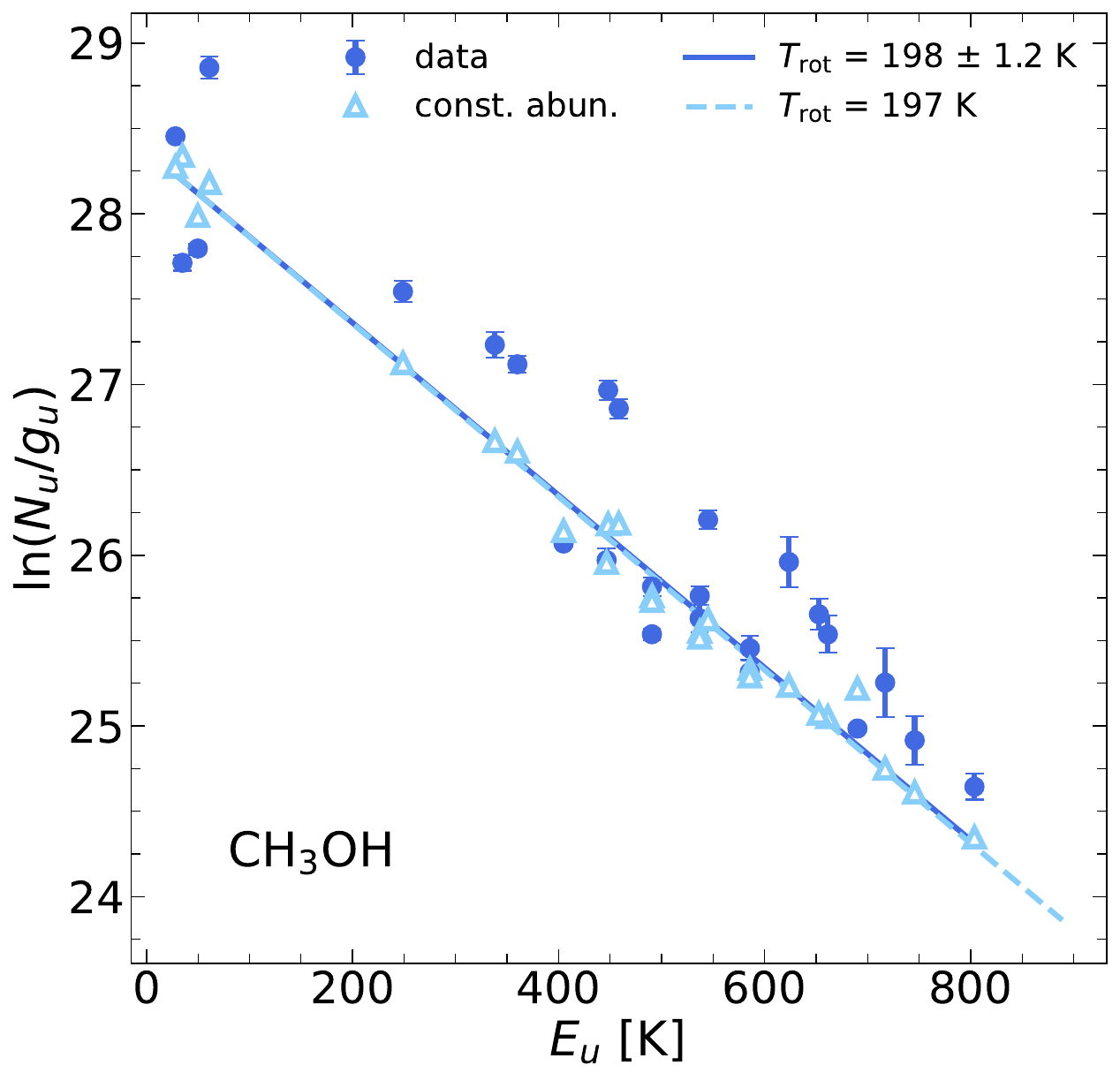}
\caption{Rotation diagram comparing 1 mm \methanol{} transitions observed in the 1 mm  NOEMA data (data: dark blue circles; fit: solid dark blue line) and a model with \methanol{} uniformly distributed within the 240 au disk. 
Simulated data for observed \methanol{} transitions (see Table \ref{tab:ch3oh}) are shown as open, light blue triangles. The fit to modeled fluxes is shown as the dashed line.
\label{fig:ch3oh}}
\end{figure}

Fig. \ref{fig:ch3oh} shows the rotation diagram for the \methanol{} observations of HOPS-370 (see also van 't Hoff et al. accepted). A fit to the data using least-squares minimization yields a rotational temperature of $T_{\rm{rot}} = 198 \pm 1.2$ K and a total column density of $N_T = (3.42 \pm 0.04)\times10^{16}$ cm$^{-2}$. Given the higher number of \methanol{} transitions with $E_u \gtrsim 400$ K relative to lower energy transitions, it is the higher energy transitions which strongly influence the derived rotational temperature. 

To model the \methanol{} emission, we adopt a constant abundance of $X(\mathrm{CH_3OH}) = 4\times10^{-8}$ distributed uniformly throughout the 240 au disk, based on the results by \cite{Tobin19, Tobin20} towards HOPS-370. This model is also shown in Fig. \ref{fig:ch3oh} and results in a rotational temperature of $T_{\rm{rot}} = 197$ K. 
Fitting the $\nu_t$ = 0 and 1 torsional states or the A- and E-symmetries separately results in very similar results. This is in agreement with the observed value.

\subsection{\methyl{}: constant abundance model} \label{subsec:ch3cn_const}
\begin{figure}[ht!]
\includegraphics[width=\columnwidth]{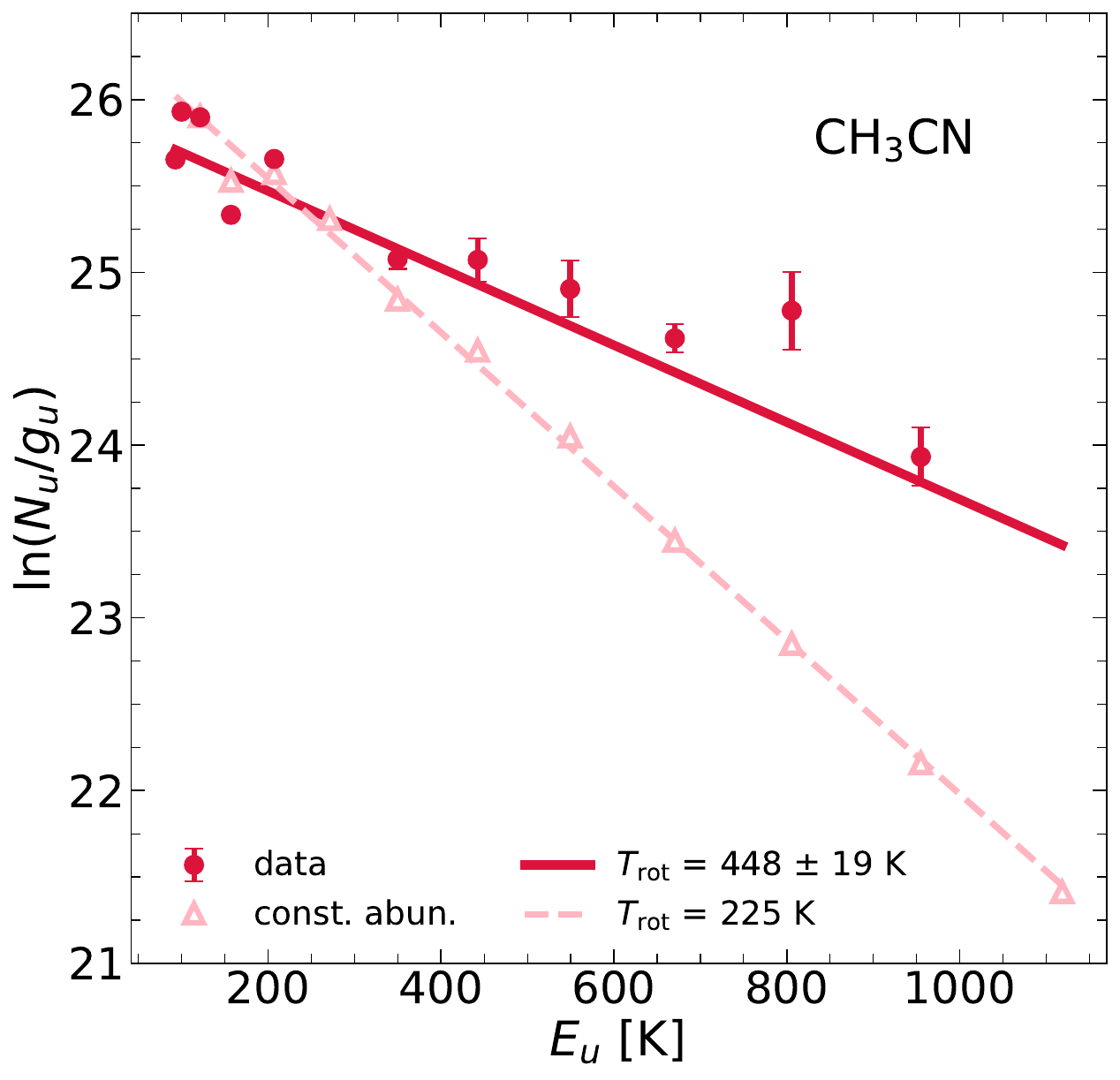}
\caption{Rotation diagram comparing 1 mm \methyl{} observations (filled crimson circles) and a model assuming a constant \methyl{} abundance of $X(\mathrm{CH_3CN}) = 5.2\times10^{-10}$ distributed uniformly throughout the entire 240 au disk (open pink triangles). The least squares fit to the observations is shown as the solid crimson line and the fit to the model is shown as the light pink dashed line.\label{fig:ch3cn_constrotdiag}}
\end{figure}

The rotation diagram for the observed K-ladder of J=14$_\mathrm{K}$-13$_\mathrm{K}$  \methyl{} transition is shown in Fig. \ref{fig:ch3cn_constrotdiag}. A least squares minimization results in an observed \methyl{} rotational temperature of $T_{\rm{rot}} = 448 \pm 19$ K and a total column density of $N_T = (8.28 \pm 0.13)\times10^{14}$ cm$^{-2}$, given the discussion in \S~\ref{subsubsec:structure}. The temperature, which is unaffected by our treatment of the partition function, is approximately 250 K higher than observed for \methanol{} and well above the estimated uncertainties. This difference likely indicates different spatial distributions for both COMs. 

To study the spatial distribution of \methyl{}, we first explore a constant abundance distributed uniformly throughout the 240 au disk, as done for \methanol{}. The \methyl{} abundance is varied to find the abundance that reproduces the observed peak K=4 flux to within 1$\sigma$ (16 mK). We choose the K=4 line as it has an intermediate upper-level energy ($E_u = 207$ K) and large signal-to-noise ($S/N \sim 70$). The K= 0 - 2 lines show some degree of blending, obfuscating the true fluxes of each line and the K=3 line (and K=6, 9, ...), while having high signal-to-noise, has a different statistical weight  due to the threefold symmetry of the methyl group and nuclear spin statistics. The K=5 transition is blended with a \methanol{} line. As such, the line K=4 has the highest signal-to-noise of unblended transitions, is unaffected by statistical weights, and has an upper energy level that is readily observable.

The rotation diagram for the full J=14-13 ladder using the resulting constant abundance $X(\mathrm{CH_3CN}) = 5.2\times10^{-10}$ is shown in Fig. \ref{fig:ch3cn_constrotdiag}. This model severely underpredicts high K fluxes. This results in a rotational temperature of only $T_{\rm{rot}} = 225$ K, which is $\sim220$ K cooler than observed. To obtain a higher rotational temperature, a higher \methyl{} abundance is required in the warmer gas closer to the star. However, simply increasing the \methyl{} abundance, results in low-K transitions being overpredicted.
Thus, a constant abundance \methyl{} distribution is unable to reproduce the NOEMA observations, in contrast to \methanol{}. 

\subsection{\methyl{}: compact abundance models} \label{subsec:ch3cn_compact}
The simplest way to increase the calculated rotational temperature while ensuring simulated fluxes agree with observations is by limiting the distribution of \methyl{} to warmer gas. In these models, we assume a constant \methyl{} abundance inside a given radius, $r_{\rm{out}}$, and zero outside. Beginning with a radius, $r_{\rm{out}} \sim 25$~au, we first systematically increase $r_{\rm{out}}$ in steps of $\sim 50$ au. Based on the effect increasing $r_{\rm{out}}$ (that is, increasing the relative amount of warm gas we are introducing) has on the $T_{\rm{rot}}$ and K=4 line flux, we refine the step size such that a better agreement between model and observations is found. At each step, we determine whether the model and observed rotation diagram agree. This is accomplished when the K=4 line flux agrees with observations to within 1$\sigma$ (16 mK) and the rotational temperature agrees to within 1$\sigma$ (19.0 K). We limit the \methyl{} abundance to be below $10^{-4}$, which corresponds to the canonical CO abundance. Table \ref{tab:compact} provides the results of a set of models that met one or both of the above listed criteria.

\begin{deluxetable*}{cccccccc} \label{tab:compact}
\tabletypesize{\scriptsize}
\tablewidth{0pt} 
\tablecaption{Summary of compact model results with varying outer extent of emission}
\tablehead{
 \colhead{$r_{\rm{out}}$} [au]& \colhead{$T_g(r_{\rm{out}})$ [K]} & \colhead{$X_{\rm{in}}(\mathrm{CH_3CN})$} & \colhead{Peak $T_{b,\mathrm{K=4}}$} & \colhead{Fits?$^{a}$} & \colhead{$T_{\rm{rot}}$ [K]} & \colhead{Fits?$^{b}$} &\colhead{$N_{T}\times10^{14}$ [$\mathrm{cm}^{-2}$]}  
}
\startdata 
110 & 193 & $6.4\times 10^{-9}$ & 1.727 & $\surd$ & 531 &         & 12.15\\ 
110 & 193 & $4.0\times 10^{-9}$ & 1.518 &         & 444 & $\surd$ & 8.56 \\ 
115 & 190 & $4.2\times 10^{-9}$ & 1.726 & $\surd$ & 436 & $\surd$ & 9.39 \\ 
120 & 187 & $5.0\times 10^{-9}$ & 1.985 &         & 452 & $\surd$ & 11.17\\ 
120 & 187 & $3.1\times 10^{-9}$ & 1.713 & $\surd$ & 386 &         & 8.02 \\ 
167 & 167 & $9.0\times 10^{-10}$ & 1.731 & $\surd$ & 257 &        & 5.13 \\ 
\enddata
\tablecomments{The observed values are: Peak $T_{b,\mathrm{K=4}} = 1.725 \pm 0.016$ K, $T_{\rm{rot}} = 448 \pm 19$, $N_{T} = (8.28 \pm 0.13)\times10^{14}$ cm$^{-2}$. }
\tablenotetext{a}{If a checkmark is present, this model reproduces the observed peak K=4 flux to within 1$\sigma$ (16 mK).}
\tablenotetext{b}{If a checkmark is present, this model reproduces the observed rotational temperature to within 1$\sigma$ (19.0 K).}
\end{deluxetable*}

\begin{figure*}[ht!]
\centering
\includegraphics[width=\textwidth]{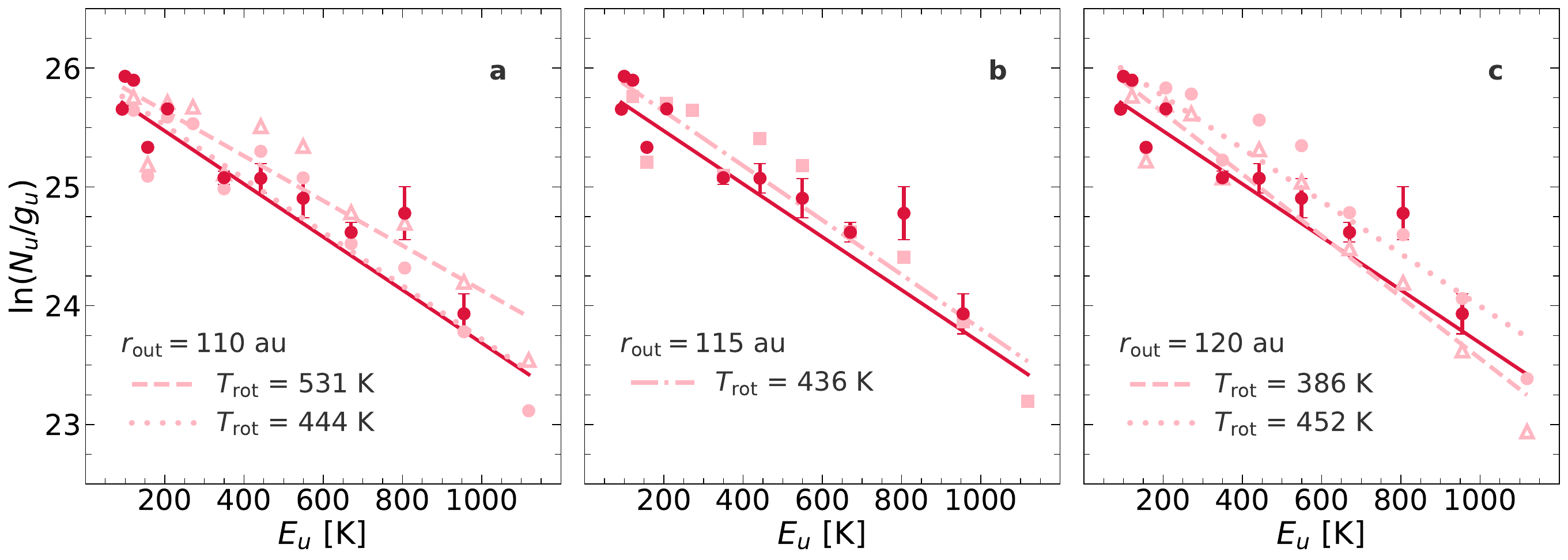}
\caption{ Rotation diagram comparing 1 mm \methyl{} observations (filled crimson circles) and models assuming a compact \methyl{} abundance distributed uniformly out to different radii, $r_{\rm{out}}$ (pink symbols). The fit to the data is shown as the solid crimson line and yields an observed rotational temperature of $T_{\rm{rot}} = 448 \pm 19$ K. Simulated data for models reproducing the observed rotational temperature and K=4 line flux to within 1$\sigma$ of their observed values are shown as filled pink squares, and the least square fit is shown as a dashed-dotted line (panel b). Simulated data for models where the rotational temperature reproduces only the observed value are shown as filled pink circles, and the fit to those data is shown as a dotted pink line (panels a and c). Simulated data for models reproducing only the K=4 line flux are shown as unfilled pink triangles, and the least squares fit to those data is shown as the light pink dashed line (panels a and c). 
\label{fig:ch3cn_comprotdiag}}
\end{figure*}

When $r_{\rm{out}} \lesssim 25$ au, the emitting area is too small, regardless of abundance; even with an abundance of $10^{-4}$, the peak K=4 flux is not reached. Systematically increasing $r_{\rm{out}}$, and thereby the emitting area, the smallest $r_{\rm{out}}$ for which the K=4 line flux can be reproduced is $\sim 100$~au. However, for these models, the rotational temperature is generally $\gtrsim 1\sigma$ larger than observed. 
A model with $r_{\rm{out}}$ = 110 au and abundance of $X(\mathrm{CH_3CN}) = 4.0\times10^{-9}$ (see Fig.~\ref{fig:ch3cn_comprotdiag}, panel a) is able to reproduce the observed rotational temperature, but yields a K=4 line flux that is $>1\sigma$ lower than observed. Increasing the abundance slightly to $X(\mathrm{CH_3CN}) = 6.4\times10^{-9}$ reproduces the K=4 flux, but results in a too high rotational temperature, as was the case for the model with $r_{\rm{out}}$ = 100 au. 
The smallest $r_{\rm{out}}$ that matches both criteria is 115~au (Fig.~\ref{fig:ch3cn_comprotdiag}, panel b). This model has an abundance of $X(\mathrm{CH_3CN}) = 4.2\times10^{-9}$.  
Thus, the inner boundary for a compact model is $\sim$110~au.

Moving $r_{\rm{out}}$ further out quickly produces results comparable to the models with $r_{\rm{out}}$ = 110 au. As shown in Fig.~\ref{fig:ch3cn_comprotdiag} (panel c), when $r_{\rm{out}}$ = 120 au, a model with $X(\mathrm{CH_3CN}) = 3.1\times10^{-9}$ is able to reproduce the observed K=4 line flux, but not the rotational temperature. On the other hand, a model with a higher abundance of $X(\mathrm{CH_3CN}) = 5.0\times10^{-9}$ reproduces the rotational temperature (to within $1\sigma$), but does not reproduce the K=4 line flux. The outer boundary for a compact model is thus $\sim$120~au. In summary, \methyl{} can be distributed as a compact abundance inside $\sim 110 - 120$ au, while the observed rotational temperature and K=4 line flux cannot be simultaneously reproduced (to within 1$\sigma$) for models outside this radial range.

\subsection{\methyl{}: jump abundance models} \label{subsec:ch3cn_jump}
The compact models illustrate the sensitivity of \methyl{} to warm gas or more strictly the need for its emission to have significant contributions from gas closer to the star.  However, these models are somewhat inconsistent with chemical expectations.  That is, nearly the entire disk lies within the \methyl{} sublimation point.  Thus \methyl{} should have an abundance that is distributed throughout the disk, albeit with a strong emission component that samples the hot gas closer to the star, but also the warm gas that lies just inside the water snowline. Another way to increase the contribution of hot gas compared to a constant or compact abundance distribution is by including a non-zero abundance outside $r_{\rm{out}}$ that is lower than the inner abundance. This is accomplished with a jump abundance model \citep{Jorgenson_2002,vanDishoeck_2006}. 

We systematically increase $r_{\rm{jump}}$ and the relative inner and outer \methyl{} abundances such that again the K=4 flux and rotational temperature are within $1\sigma$ of observed values.  As before, we only consider abundances less than the canonical CO abundance; that is, $X_{\rm{in}}(\mathrm{CH_3CN}) < 10^{-4}$. 
For realistic lower limits on the \methyl{} abundance in the outer disk, we adopt an abundance ratio $X(\mathrm{CH_3CN})/X(\mathrm{CH_3OH})$ of 0.1\% based on the weighted-average column density ratio presented by \citet{Nazari_2022} using observations of high-mass protostars in the ALMAGAL sample and literature values toward low- and intermediate-mass protostars. This ratio is $\sim$10 times lower than found by, for example, \citep{Belloche_2020} (CALYPSO survey) and \citet{Yang_2021} (PEACHES survey), so the lowest outer abundance we consider ($X_{\rm{out}}(\mathrm{CH_3CN}) \sim 10^{-11}$) may be conservative.

\begin{figure*}[ht!]
\centering
\includegraphics[width=0.95\textwidth]{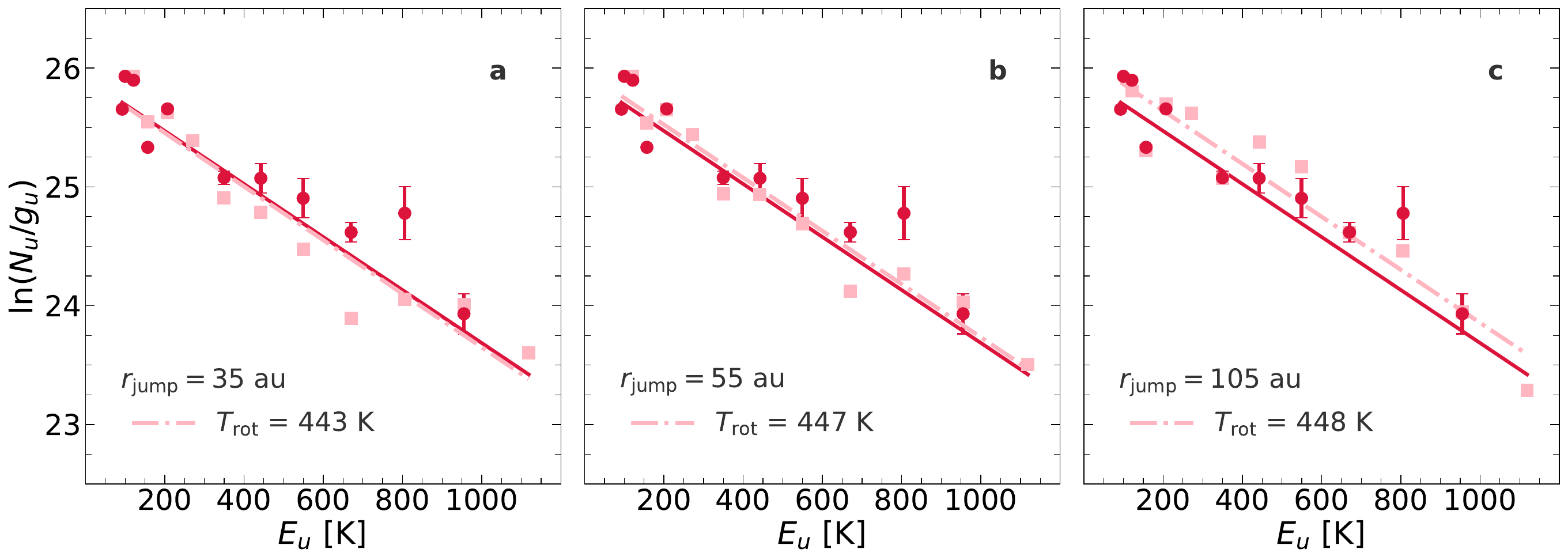}
\caption{Rotation diagram comparing 1 mm \methyl{} observations (filled crimson circles) and jump-abundance models assuming non-zero \methyl{} abundances distributed uniformly inside and outside varies jump radii, $r_{\rm{jump}}$ (pink squares). The fit to the data is shown as the solid crimson line and yields an observed rotational temperature of $T_{\rm{rot}} = 448 \pm 19$ K. The fits to the model data are shown as the dotted-dashed pink line and result the rotational temperatures listed in each panel. See Table \ref{tab:jump} for a summary of the jump-abundance model results. \label{fig:ch3cn_jumpcombrotdiag}}
\end{figure*}

Table \ref{tab:jump} shows the result of varying the jump location and corresponding inner/outer abundances required to reproduce the observations. In each case, there is a \methyl{} enhancement of $\gtrsim 15$ inside the jump radius relative to outside. Fig. \ref{fig:ch3cn_jumpcombrotdiag} shows rotation diagrams for a range of these models.

\begin{deluxetable*}{ccccccc} \label{tab:jump}
\tabletypesize{\scriptsize}
\tablewidth{0pt} 
%\tablenum{1}
\tablecaption{Summary of jump-abundance models that can reproduce the observations.}
\tablehead{
 \colhead{$r_{\rm{jump}}$ [au]} & \colhead{$T_g(r_{\rm{jump}})$ [K]} & \colhead{$X_{\rm{in}}(\mathrm{CH_3CN})$} & \colhead{$X_{\rm{out}}(\mathrm{CH_3CN})$} & \colhead{Peak $T_{b,\mathrm{K=4}}$ [K]} & \colhead{$T_{\rm{rot}}$ [K]} & \colhead{$N_{T}\times10^{14}\ [\mathrm{cm}^{-2}]$}
}
\startdata 
35  & 288 & $9.60\times 10^{-7 }$ & $4.90\times 10^{-10}$ & 1.721 & 444 & 8.00 \\
55  & 246 & $1.90\times 10^{-8 }$ & $4.60\times 10^{-10}$ & 1.727 & 447 & 8.64 \\
90  & 207 & $6.10\times 10^{-9 }$ & $2.70\times 10^{-10}$ & 1.726 & 445 & 9.66 \\
100 & 199 & $5.30\times 10^{-9 }$ & $1.70\times 10^{-10}$ & 1.712 & 448 & 9.70 \\
105 & 196 & $4.95\times 10^{-9 }$ & $1.15\times 10^{-10}$ & 1.715 & 448 & 9.72 \\
\enddata
\tablecomments{The observed values are: Peak $T_{b,\mathrm{K=4}} = 1.725 \pm 0.016$ K, $T_{\rm{rot}} = 448 \pm 19$, $N_{T} = (8.28 \pm 0.13)\times10^{14}$ cm$^{-2}$. }
\end{deluxetable*}

While not shown in Fig. \ref{fig:ch3cn_jumpcombrotdiag}, when the jump location is located inside 35 au (where the disk temperature is $\gtrsim 290$ K), levels with K$\leq4$ are generally well-populated, while higher K-levels are not. For $K \geq 6$, peak line fluxes are $\sim45\%$ lower than observed on average. This ultimately results in a derived rotational temperature about 200 K lower than observed, even with an inner \methyl{} abundance in excess of the canonical CO abundance.
This is primarily due to having too small an emitting area as well as optically thick dust obscuring the emission from the inner most region.  However, a jump at 35 au with an abundances of $X_{\rm{in}}(\mathrm{CH_3CN}) = 9.60\times 10^{-7 }$  and $X_{\rm{out}}(\mathrm{CH_3CN}) = 4.90\times 10^{-10}$ provides a good match to the \methyl{} observations. 

Good agreement between model and observation is found for models with jump radii out to 105 au (Fig. \ref{fig:ch3cn_jumpcombrotdiag} and Table \ref{tab:jump}). As $r_{\rm{jump}}$ is increased we also see that while inner/outer abundances decrease (as expected and also seen in the compact abundance distributions), the inner abundance is consistently $\gtrsim15$ larger than the outer abundance.  Furthermore, we see that despite retrieving the observed rotational temperature and K=4 peak flux, the total column density increases as the jump radius is moved outward. This is expected since more material is being added as the radius increases. Our most concentrated jump model, with $r_{\rm{jump}} = 35$~au has a total column $\sim2\sigma$ lower than observed. In contrast, our best-fit compact abundance model with $r_{\rm{out}}=115$~au, results in a total column that is $\gtrsim 11\sigma$ larger than observed. 
This is to say, the observed total column necessitates a jump abundance model with \methyl{} being distributed inside $\sim$55 au, much more concentrated than the best-fit compact abundance model. 

When the jump location is shifted outward of 105 au (where the disk gas temperature is $<196$ K), the jump model stops being as effective as, for example, the compact emission model with $r_{\rm{out}}=110$ au at reproducing observed values. Outside 105 au, the contribution from the warm rather than hot gas begins to dominate, and it becomes difficult to reproduce emission from the high K-levels. 

To summarize, models with $35 < r_{\rm{jump}} < 105$ au are able to simultaneously reproduce the observed rotational temperature as well as the peak K=4 flux. For models with $r_{\rm{jump}} < 35$ au, the emitting area is too small, and, as such, the rotational temperature cannot be reproduced despite using an abundance of $10^{-4}$. For models where $r_{\rm{jump}}$ is larger than 105 au, we cannot distinguish whether \methyl{} is distributed as a jump model or compact abundance. Furthermore, when the total column density is considered, models indicate \methyl{} needs to have a compact component with the jump radius being $r_{\rm{jump}} \lesssim 55$~au.

\section{Discussion} \label{sec:discussion}

We observe \methyl{} to have a rotational temperature of $T_{\rm{rot}} = 448 \pm 19$ K, which is approximately 250 K warmer than observed for \methanol{}. Models of their spatial distribution indicate that \methyl{} must indeed be different from \methanol{}, with \methyl{} shown to be enhanced in the warm gas relative to \methanol{}. Nominally this requires a \methyl{} production mechanism in the hot gas or a destruction mechanism preferentially diminishing \methyl{} in the warm gas. Below we explore how detailed excitation might influence our results and discuss potential mechanisms that could enhance \methyl{} in the hot gas or deplete it in the warm gas.

\subsection{Effect of temperature profile choice} \label{subsec:iso}
\begin{figure}[ht!]
    \centering
    \includegraphics[width=\columnwidth]{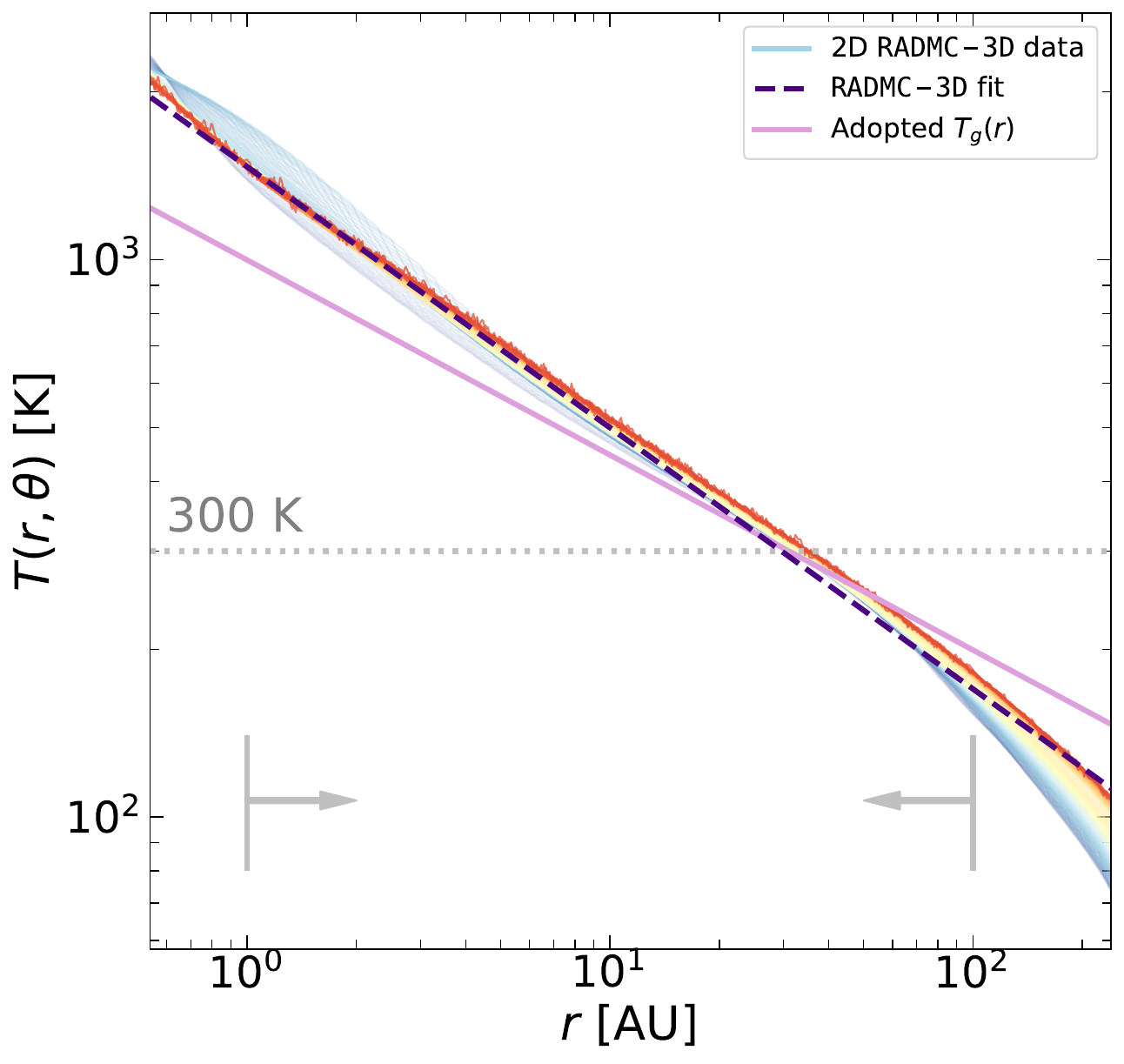}
    \caption{Comparison between the 2D temperature profile in the disk and inner envelope for the best-fit model from \citet{Tobin20} calculated with \texttt{RADMC-3D} (color gradient, where red indicates $\theta \rightarrow 0$ radians, and blue indicates $\theta \rightarrow \pi/2$ radians, i.e. approaching the midplane) and the power law temperature profile adopted in this work based on modeling by \citet{Tobin20} (pink line). The black dashed line represents a power-law fit to the \texttt{RADMC-3D} temperature profile within the region demarcated by the gray lines ($T_g(r) \approx 1470\mathrm{K}\ (r/\mathrm{au})^{-0.47}$). \label{fig:iso}}
    \end{figure}

The choice of temperature structure could have a significant impact on the outcome of our modelling work \citep[see e.g.,][ and references therein]{Calahan_2021}. A potentially influential choice in our model is the assumption of a vertically isothermal temperature profile. \cite{Aikawa_2002} and \cite{Dartois_2003} illustrated that the temperature structure for an evolved disk is not simply vertically isothermal. However, for deeply embedded protostars, the addition of an envelope (and its opacity) acts to intercept and reprocess the incident starlight \citep[e.g.,][]{DAlessio_1999, Tobin19}. Ultimately, this results in only a weak vertical temperature gradient compared to more evolved disks. This is confirmed for HOPS-370 specifically by 2D \texttt{RADMC-3D} calculations \citep{Tobin20}, as shown in Fig.~\ref{fig:iso}; the temperature does not vary greatly as a function of polar angle (height), especially between 1 and 100 au (demarcated by the gray arrows). In other words, the assumption of a vertically isothermal structure is reasonable for HOPS-370.

Assuming the temperature is vertically isothermal implies that a 1D temperature profile can be used to describe the entire 2D disk temperature. We adopt the power-law temperature profile from \citet{Tobin20}, who model the molecular line emission and dust continuum, albeit separately. This is done because calculating the dust temperature self-consistently using the radiative transfer calculations of \texttt{RADMC-3D} is computationally expensive. This quickly becomes intractable when full dust thermal calculations are to be done for each model with a different spatial distribution for multiple molecular species. To assess the validity of this simplified temperature profile, Fig. \ref{fig:iso} shows the difference between the dust temperature calculated self-consistently with \texttt{RADMC-3D} and the power-law temperature profile adopted in \S~\ref{subsubsec:structure} based on the NOEMA \methanol{} observations.

A power-law fit to the 2D \texttt{RADMC-3D} data yields:
\begin{equation}
T_{g+d}(r) = 1470\mathrm{K}\ \Big(\frac{r}{\mathrm{au}} \Big)^{-0.47}
\end{equation}
whereas in \S~\ref{subsec:ch3oh} we find that the observed \methanol{} rotation diagram can be described with the following power law as modelled by \cite{Tobin20}:
\begin{equation} \label{eq:John_mod}
T_{g}(r) = 998.7\mathrm{K}\ \Big(\frac{r}{\mathrm{au}}\Big)^{-0.35}.
\end{equation}
With a larger $T_0$ and steeper power-law index, $q$, the fit to the \texttt{RADMC-3D} calculated temperature profile is hotter at 1 au but colder further out in the disk ($>$ 10 au) than the power-law we assume (i.e. Eq.~\ref{eq:John_mod}). This could affect our constraint on the jump location. However, even at 240 au, where there is the largest deviation between the two temperature profiles, we see there is at most a difference of $\sim35$ K between models. Any difference between the temperature structures where the jumps were placed is less than a factor of 2, which is less than $\sim50$ K. Even though a slightly different radial temperature profile may influence the exact location of the outer radius in the compact models or the jump radius in the jump-abundance models, it will not affect the conclusion that \methyl{} is differently distributed than \methanol{} within the snowline of HOPS-370.

\subsection{Pebble drift}  \label{subsec:pebbles}
One potential explanation for the existence of an increase in \methyl{} in the inner disk could be pebble drift. From the perspective of dynamics, dust of different sizes interacts differently with the gas. The small dust is dynamically well-coupled to the gas, while large dust experiences maximum inward radial drift. The presence of a jump abundance distribution could in principle be explained by the inward drift of large dust, causing an enhanced abundance in the inner parts of the disk where the ice sublimates \citep[e.g.,][]{Krijt_2018, Zhang_2020a}.

However, if radial drift were occurring and giving rise to the jump abundance seen in \methyl{}, we would also expect to a jump abundance distribution for \methanol{} unless there is some additional ice chemistry serving to keep \methanol{} enhanced outside the \methyl{} outer radius or jump location in the outer disk. Since the simpler constant abundance model reproduces the \methanol{} fluxes observed by NOEMA, a jump model is unnecessary. Thus, radial drift is unlikely to be the reason for the difference in abundance distributions between \methanol{} and \methyl{}.

\subsection{Destroying CH$_3$CN in warm gas} \label{subsec:destroy}
Models of their spatial distribution indicate inner \methyl{} abundances are heightened by a factor of $\sim 15$ relative to outer within $\sim105$~au, where warm gas ($150\ \mathrm{K} \lesssim T_g \lesssim 300$ K) resides. This enhancement could be due to a chemical mechanism reducing the \methyl{} in the warm gas. 

Two such destruction mechanisms could be gas-phase reactions between He$^+$ or H$_3^+$ for both \methyl{} and \methanol{}. The reaction between  He$^+$ and \methyl{} occurs on a timescale similar to that for He$^+$ and \methanol{}.  However,  the reaction between  H$_3^+$ and \methyl{} is approximately a factor of three faster than for \methanol{} \citep{McElroy_2013}. After ice sublimation this would lead to quicker re-equilibration towards chemical equilibrium for \methyl{} with a timescale that is a factor of three shorter.   This could lead to a shift in the abundance of \methyl{} relative to \methanol{} in the outer disk gas, but any model would have to be fine-tuned in timing to capture this change within the short period of time as timescales are $< 10^5$~yrs at these high densities \citep{Hatchell98}.
Thus, the elevated \methyl{} abundance is difficult to explain by a gas-phase destruction mechanism.

\subsection{Carbon-grain sublimation} \label{subsec:subl}
We see \methyl{} is characterized by hot gas, where there is a relatively concentrated \methyl{} abundance enhanced relative to \methanol{}. This could indicate the presence of hot gas phase chemistry taking place, which could be the result of carbon grain sublimation. Carbon-grain sublimation is the process by which molecules in refractory carbon grains are released to the gas. This occurs at $\gtrsim 300$ K; the so-called ``soot line" \citep{Kress_2010}.

When dust grains are destroyed at the soot line, more carbon and nitrogen will be observed in the hot gas relative to oxygen, since the cometary carbon and nitrogen budget is higher for dust grains than for ices, while the reverse is true for the cometary oxygen budget \citep{Rubin_2019}.  After the initial release of carbon- and nitrogen-bearing species in the hot gas, \methyl{} could then form there via reactions between CH$_3^{+}$ and HCN, the product of which, CH$_3$CNH$^+$, rapidly equilibrates to \methyl{} via electronic dissociative recombination \citep{Garrod_2022}. The conditions in HOPS-370 could facilitate these gas phase reactions.

\cite{Merel_2020} provide three potential signatures of carbon grain sublimation and top-down chemistry to occur within young disks: 
\begin{enumerate}
\item An increased abundance of  N-COMs e.g., \methyl{} inside the soot line,
\item N-COMs being distributed within a smaller spatial area than O-COMs e.g., \methanol{}, and therefore
\item N-COMs exhibiting higher excitation temperatures than O-COMs. 
\end{enumerate}
Potential evidence for carbon-grain sublimation has been observed by \cite{Nazari_2023} for high mass protostars in the ALMAGAL survey. However, we also see potential evidence in HOPS-370.

The excitation temperature for the 1 mm NOEMA observations of \methyl{} (see also van 't Hoff et al. accepted) is $T_{\rm{rot}} = 448 \pm 19$ K, approximately 250 K warmer than observed for \methanol{} ($T_{\rm{rot}} = 198 \pm 1.2$ K). 

While we find a compact \methyl{} abundance out to 115 au is able to reproduce the observations, so do several jump abundance models. All of these jump abundance models have the jump occurring between $35 \lesssim r_{\rm{jump}} \lesssim 105$ au, corresponding to $196 < T_g < 288$ K. Models with smaller jump radii (e.g. $r_{\rm{jump}} \lesssim 55$~au) are favored based on the total column density calculated being most similar to observations. 
While the 300 K line is expected to correspond to the soot line, our analysis is agnostic of this.  However, we clearly see an enhanced \methyl{} abundance inside of the expected soot line location. Thus, criteria 1 and 3 are satisfied.

The rotation diagram analysis we perform significantly constrains the location where any hot gas chemistry might be taking place, despite the size of the NOEMA beam. Regardless, future observations with high angular resolution could spatially resolve the location of the water snowline ($\sim225$~au or $0\farcs95$) and maybe even the putative soot line ($\sim31$~au or $0\farcs13$ assuming the soot line occurs at 300~K for HOPS-370).   Regardless, there is clear evidence for chemical shifts in this system that hint at changes that can be resolved.

\section{Conclusion} \label{sec:conclusion}
With spatially-unresolved 1 mm NOEMA observations of \methanol{} and \methyl{} in the protostellar disk around HOPS-370, 
we aim to determine if \methanol{} and \methyl{} are distributed differently. 
We use the radiative transfer modelling software \texttt{LIME} in combination with a source-specific physical model to calculate spectra for each of these molecules for different parameterized spatial distributions. Rotation diagrams are used to analyze the observed and modeled spectra. Our main findings are as follows:
\begin{itemize}
\item \methanol{} has an observed rotational temperature of $T_{\rm{rot}} = 198 \pm 1.2$ K, while \methyl{} is observed to be significantly warmer with $T_{\rm{rot}} = 448 \pm 19$ K.
\item The \methanol{} observations can be reproduced with a constant abundance model out to 240 au.
\item The \methyl{} observations cannot be reproduced by a uniform distribution similar to \methanol{}. Instead, the \methyl{} emission is either described by a more compact abundance distribution within $\sim115$~au or by a jump-abundance distribution with the increase in abundance happening for radii between $\sim$35 and $\sim$105 au. In the inner region, the \methyl{} is enhanced by a factor of $\sim15$ compared to the outer region.  

\end{itemize}

\noindent These changes suggest chemical gradients are present inside the water snowline within hot gas.
HOPS-370 is a unique laboratory for characterization of chemistry in hot gas which is needed to understand the formation of inner solar/stellar system planets like our own.  

\acknowledgments 
L.G.W. acknowledges support from the Rackham Merit Fellowship, under which a majority of this work was funded. M.L.R.H. acknowledges support from the Michigan Society of Fellows.

\appendix
\section{Molecular data}
\begin{figure}[ht!]
    \centering
    \includegraphics[width=\textwidth]{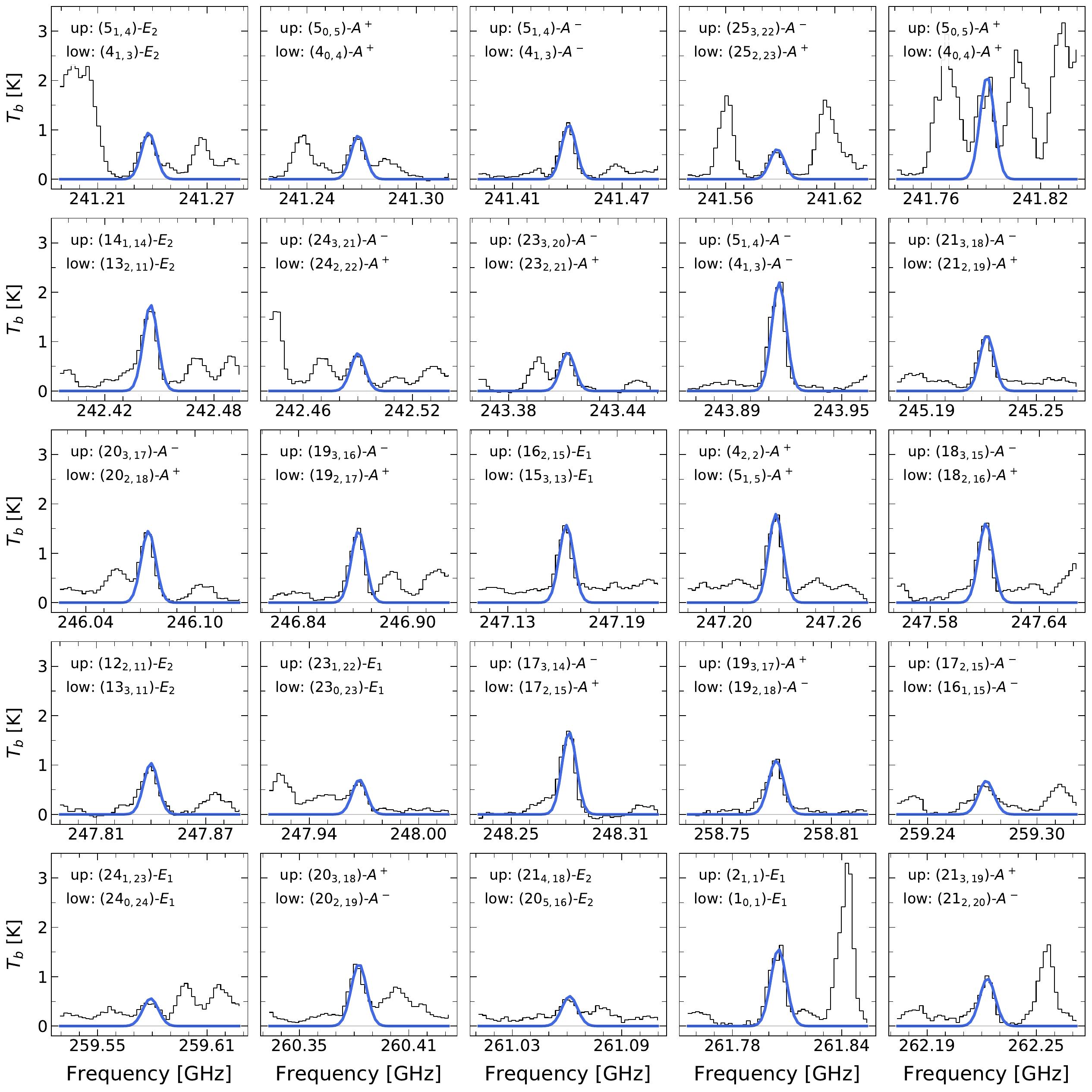}
    \caption{Gaussian fits (blue) to observed \methanol{} transitions (see also Table \ref{tab:ch3oh}). The lines are sorted according to transition frequency in GHz, with each panel centered on the line. The source velocity is $v_{LSR} = 11.2$ km s$^{-1}$ and each line has a full-width half-maximum of 11.5 km s$^{-1}$.
    \label{fig:ch3oh_spectra} }
    \end{figure}

Tables~\ref{tab:ch3oh} and \ref{tab:ch3cn} provide spectroscopic data for observed and modelled transitions of \methanol{} and \methyl{}, respectively (see also \S~\ref{subsubsec:structure}). For \methanol{}, we provide in Fig. \ref{fig:ch3oh_spectra} the observed transitions (with quantum numbers as they appear in the \texttt{CDMS} database), along with the resultant Gaussian fits calculated with \texttt{GILDAS}.

\begin{deluxetable}{cccccccc} \label{tab:ch3oh}
\tabletypesize{\scriptsize}
\tablewidth{0pt} 
\tablecaption{Spectroscopic data for \methanol{} obtained from \texttt{CDMS}}
\tablehead{\colhead{Upper level} & \colhead{Lower level} & \colhead{Frequency} & \colhead{$E_u$} & \colhead{$g_u$} & \colhead{$A_{ul}$} & \colhead{Observed flux} & \colhead{Flux error} \\ 
\colhead{$(J_{K_a, K_c})$}  & \colhead{$(J_{K_a, K_c})$}  &\colhead{[MHz]} & \colhead{[K]} & \colhead{} & \colhead{[s$^{-1}$]} & \colhead{[K km s$^{-1}$]} & \colhead{[K km s$^{-1}$]} \\ 
}
\startdata
\hline
 & & & $v_t = 0$ & &  & & \\
\hline
% A
$(25_{3, 22})$ $A^-$ & $(25_{2, 23})$ $A^+$ & 241588.758 & 803.7 & 204 & 8.19e-05 & 7.4304 & 0.56  \\ 
$( 5_{0,  5})$ $A^+$ & $( 4_{0,  4})$ $A^+$ & 241791.352 &  34.8 &  44 & 6.05e-05 & 25.436 & 1.17  \\ 
$(24_{3, 21})$ $A^-$ & $(24_{2, 22})$ $A^+$ & 242490.245 & 745.7 & 196 & 8.19e-05 & 9.3002 & 1.33  \\ 
$(23_{3, 20})$ $A^-$ & $(23_{2, 21})$ $A^+$ & 243412.610 & 690.1 & 188 & 8.20e-05 &  9.499 & 0.193 \\ 
$( 5_{1,  4})$ $A^-$ & $( 4_{1,  3})$ $A^-$ & 243915.788 &  49.7 &  44 & 5.97e-05 & 26.813 & 0.714 \\ 
$(21_{3, 18})$ $A^-$ & $(21_{2, 19})$ $A^+$ & 245223.019 & 585.7 & 172 & 8.23e-05 & 13.751 & 0.977 \\ 
$(20_{3, 17})$ $A^-$ & $(20_{2, 18})$ $A^+$ & 246074.605 & 537.0 & 164 & 8.25e-05 & 17.744 & 0.978 \\ 
$(19_{3, 16})$ $A^-$ & $(19_{2, 17})$ $A^+$ & 246873.301 & 490.7 & 156 & 8.27e-05 & 17.737 & 0.979 \\ 
$( 4_{2,  2})$ $A^+$ & $( 5_{1,  5})$ $A^+$ & 247228.587 &  60.9 &  36 & 2.12e-05 & 21.876 & 1.44  \\ 
$(18_{3, 15})$ $A^-$ & $(18_{2, 16})$ $A^+$ & 247610.918 & 446.6 & 148 & 8.29e-05 & 19.583 & 1.39  \\ 
$(17_{3, 14})$ $A^-$ & $(17_{2, 15})$ $A^+$ & 248282.424 & 404.8 & 140 & 8.30e-05 & 20.338 & 0.258 \\ 
$(19_{3, 17})$ $A^+$ & $(19_{2, 18})$ $A^-$ & 258780.248 & 490.6 & 156 & 9.01e-05 & 13.311 & 0.451 \\ 
$(20_{3, 18})$ $A^+$ & $(20_{2, 19})$ $A^-$ & 260381.463 & 536.9 & 164 & 9.14e-05 & 15.384 & 1.22  \\ 
$(21_{3, 19})$ $A^+$ & $(21_{2, 20})$ $A^-$ & 262223.872 & 585.6 & 172 & 9.28e-05 & 11.785 & 0.879 \\ 
 \hline
%E
$(14_{1, 14})$ $E_2$ & $(13_{2, 11})$ $E_2$ & 242446.084 & 248.9 & 116 & 2.29e-05 & 21.325 & 1.32  \\ 
$(16_{2, 15})$ $E_1$ & $(15_{3, 13})$ $E_1$ & 247161.950 & 338.1 & 132 & 2.57e-05 & 19.189 & 1.45  \\ 
$(23_{1, 22})$ $E_1$ & $(23_{0, 23})$ $E_1$ & 247968.119 & 661.4 & 188 & 4.44e-05 & 8.6042 & 0.934 \\ 
$(24_{1, 23})$ $E_1$ & $(24_{0, 24})$ $E_1$ & 259581.398 & 717.0 & 196 & 4.91e-05 & 6.8224 & 1.38  \\ 
$(21_{4, 18})$ $E_2$ & $(20_{5, 16})$ $E_2$ & 261061.320 & 623.6 & 172 & 3.02e-05 & 7.3807 & 1.09  \\ 
$( 2_{1,  1})$ $E_1$ & $( 1_{0,  1})$ $E_1$ & 261805.675 &  28.0 &  20 & 5.57e-05 & 19.046 & 0.315 \\ 
\hline
 & & & $v_t = 1$ & &  & & \\
\hline
% A
$( 5_{0,  5})$ $A^+$ & $( 4_{0,  4})$ $A^+$ & 241267.862 & 458.4 &  44 & 6.00e-05 & 10.786 & 0.621 \\
$( 5_{1,  4})$ $A^-$ & $( 4_{1,  3})$ $A^-$ & 241441.270 & 360.0 &  44 & 5.79e-05 & 13.471 & 0.643 \\
$(17_{2, 15})$ $A^-$ & $(16_{1, 15})$ $A^-$ & 259273.686 & 652.7 & 140 & 5.58e-05 & 8.2852 & 0.74  \\
 \hline
%E
$( 5_{1,  4})$ $E_2$ & $( 4_{1,  3})$ $E_2$ & 241238.144 & 448.1 &  44 & 5.75e-05 & 11.515 & 0.654 \\ 
$(12_{2, 11})$ $E_2$ & $(13_{3, 11})$ $E_2$ & 247840.050 & 545.1 & 100 & 6.28e-05 & 12.673 & 0.696 \\ 
\hline
\enddata
\tablecomments{The quantum numbers (QNs) for each level of the transition are listed as $(J_{K_a, K_c})$, where $J$ is the total rotational angular momentum, $K_a$ is the projection of $J$ onto the $A$ and $C$ inertial axes, respectively. Following the parentheses in each entry is the symmetry $A$ or $E$. For transitions with $A$-symmetry, the parity is denoted with + or -. For transitions with $E$-symmetry, the subscript 1 indicates $K_a \geq 0$, while the subscript 2 indicates $K_a < 0$} 
\end{deluxetable}

\begin{deluxetable}{cccrrccc} \label{tab:ch3cn}
\tabletypesize{\scriptsize}
\tablewidth{0pt} 
\tablecaption{Spectroscopic data for \methyl{} obtained from LAMDA}
\tablehead{\colhead{Upper level} & \colhead{Lower level} & \colhead{Frequency} & \colhead{$E_u$} & \colhead{$g_u^a$} & \colhead{$A_{ul}$ } & \colhead{Observed flux } & \colhead{Flux error } \\
\colhead{$J_K$} & \colhead{$J_K$} & \colhead{[MHz]} & \colhead{[K]} & \colhead{} & \colhead{[s$^{-1}$]} & \colhead{[K km s$^{-1}$]} & \colhead{[K km s$^{-1}$]}
}
\startdata
\hline
 & & & & $v_t = 0$ &  & &\\
\hline
$14_{12}$ & $13_{12}$ & 256817.163 & 1118.7 & 29.0 & 3.89e-4 & ---    & ---   \\ 
$14_{11}$ & $13_{11}$ & 256930.140 &  955.2 & 14.5 & 5.62e-4 & 1.5713 & 0.266 \\ 
$14_{10}$ & $13_{10}$ & 257033.444 &  805.8 & 14.5 & 7.20e-4 & 4.6887 & 1.05  \\ 
$14_{9} $ & $13_{9}$  & 257127.036 &  670.5 & 29.0 & 8.63e-4 & 9.5756 & 0.78  \\ 
$14_{8} $ & $13_{8}$  & 257210.878 &  549.4 & 14.5 & 9.91e-4 & 7.317  & 1.2   \\ 
$14_{7} $ & $13_{7}$  & 257284.935 &  442.4 & 14.5 & 1.11e-3 & 9.6442 & 1.2   \\ 
$14_{6} $ & $13_{6}$  & 257349.180 &  349.7 & 29.0 & 1.20e-3 & 21.106 & 1.2   \\ 
$14_{5} $ & $13_{5}$  & 257403.585 &  271.2 & 14.5 & 1.29e-3 & ---    & ---   \\ 
$14_{4} $ & $13_{4}$  & 257448.128 &  207.0 & 14.5 & 1.36e-3 & 21.196 & 0.301 \\ 
$14_{3} $ & $13_{3}$  & 257482.792 &  157.0 & 29.0 & 1.41e-3 & 31.866 & 0.269 \\ 
$14_{2} $ & $13_{2}$  & 257507.562 &  121.3 & 14.5 & 1.45e-3 & 28.774 & 0.305 \\ 
$14_{1} $ & $13_{1}$  & 257522.428 &   99.8 & 14.5 & 1.47e-3 & 30.202 & 0.198 \\ 
$14_{0} $ & $13_{0}$  & 257527.384 &   92.7 & 14.5 & 1.48e-3 & 23.013 & 0.338 \\ 
\enddata
\tablenotetext{a}{The statistical weights presented in the \texttt{CDMS} entry for \methyl{} have the contribution from the nuclear spin factored out. However, those present in the \texttt{LAMDA} file (and used in our calculation of the total column density) have the nuclear spin factor included. See, for example, \cite{Townes_1975} and \cite{Mangum_2015} for discussion of the statistical weights for molecules with $C_{3\nu}$ symmetry, like \methyl{}.}
\tablecomments{While not used in constructing the observed rotation diagram, we include the K=5 and 12 lines in our models and the linear regressions. In both observed and model rotation diagrams, we do not include the K=0 and 1 lines in the rotation diagram analysis as these transitions are blended.} 
\end{deluxetable}


\vspace{5mm}
\facilities{NOEMA}


\software{\texttt{RADEX} \citep{radex},
        \texttt{GILDAS} \citep{Pety_2005, GILDAS_2013}, 
        \texttt{LIME} \citep{lime},
        \texttt{RADMC-3D} \citep{radmc3d},
        \texttt{optool}, \citep{optool},
        astropy \citep{astropy:2013}, \citep{astropy:2018, astropy:2022}, 
        SciPy \citep{scipy},
        Matplotlib \citep{plt}}

%% the bibliography. The sample63.bib file is populated from ADS. To
\bibliography{warmgas}{}
\bibliographystyle{aasjournal}



\end{document}